\documentclass[%
 reprint,
superscriptaddress,
 amsmath,amssymb,
 aps,
prb,
]{revtex4-2}

\usepackage{graphicx}
\usepackage{dcolumn}
\usepackage{bm}

\usepackage{orcidlink}

\begin{document}

\preprint{APS/123-QED}

\title{Giant enhancement of the transverse magneto-optical Kerr effect in etchless bismuth-substituted yttrium iron garnet empowered by quasi-bound states in the continuum}

\author{Qin Tang}
\affiliation{School of Physics and Materials Science, Nanchang University, Nanchang 330031, China}

\author{Dandan Zhang}
\affiliation{School of Physics and Materials Science, Nanchang University, Nanchang 330031, China}

\author{Shuyuan Xiao~\orcidlink{0000-0002-4446-6967}}%
\email{syxiao@ncu.edu.cn}
\affiliation{School of Information Engineering,
Nanchang University, Nanchang 330031, China}
\affiliation{Institute for Advanced Study, Nanchang University, Nanchang 330031, China}

\author{Meibao Qin}%
\affiliation{School of Education, Nanchang Institute of Science and Technology, Nanchang 330108, China.}

\author{Jizhou He}%
\affiliation{School of Physics and Materials Science, Nanchang University, Nanchang 330031, China}

\author{Tingting Liu~\orcidlink{0000-0003-3671-2782}}%
\affiliation{School of Information Engineering,
Nanchang University, Nanchang 330031, China}
\affiliation{Institute for Advanced Study, Nanchang University, Nanchang 330031, China}

\author{Qinghua Liao}%
\affiliation{School of Physics and Materials Science, Nanchang University, Nanchang 330031, China}

\author{Tianbao Yu~\orcidlink{0000-0003-4308-7022}}%
\email{yutianbao@ncu.edu.cn}
\affiliation{School of Physics and Materials Science, Nanchang University, Nanchang 330031, China}%

\date{\today}

\begin{abstract}
Here, we propose an etchless bismuth-substituted yttrium iron garnet layer assisted by a one-dimensional resonant grating waveguide to enhance transverse magneto-optical Kerr effect (TMOKE) via the excitation of quasi-bound state in the continuum. The TMOKE amplitude can be tailored by manipulating the perturbation parameter, and it can reach as high as 1.978, approaching the theoretical maximum value of 2. Additionally, a single-mode temporal coupled-mode theory is employed to further reveal the underlying physical mechanism. It is found that TMOKE is strongly related to the line width of the quasi-BIC resonance and local field enhancement, which are pivotal factors in the design and optimization of photonic devices. As a potential application, we design and numerically demonstrate a refractive index sensor based on the resonantly enhanced TMOKE, with the optimal sensitivity of 110.66 nm/RIU and the corresponding maximum figure of merit of 299.3 RIU$^{-1}$. Our work provides a simple and efficient approach for enhancing TMOKE based on an easy-to-fabricate platform, laying the groundwork for exploring and developing magneto-optical devices such as sensors, magnetic storage devices, and nonreciprocal photonic devices. 

\end{abstract}

\maketitle

\section{Introduction}
The transverse magneto-optical Kerr effect (TMOKE), typically characterized by the relative change in the intensity of the reflected light, has important applications in the fields such as biosensing \cite{1}, magnetometry \cite{2}, and ultrafast magnetism \cite{3,4}. It is challenging to observe TMOKE on smooth magneto-optical thin films due to the weak light-matter interaction. To address this dilemma, there have been many attempts to enhance TMOKE through the integration of dielectric/plasmonic nanostructures with magneto-optical thin films, especially by utilizing resonance phenomena such as localized surface plasmon resonance/surface plasmon resonance \cite{5,6,7,8,9,10,11,12,13,14,15}, guided mode resonance \cite{16,17,18}, and Mie resonance \cite{19}. However, these structures are constrained by absorption and scattering losses to exhibit low-quality-factor ($Q$-factor) magneto-optical responses, thereby limiting the applicability in more delicate applications.

In recent years, bound states in the continnum (BICs) have provided a promising avenue to enhance the interactions between light and magneto-optical material due to their unique capability to confine energy \cite{20,21,22,23,24}. BICs are radiationless eigenstates embedded to the radiation continuum, which can be viewed as resonances with vanishing linewidths and infinite $Q$-factors \cite{25,26}. However, for real systems, BICs can transform into quasi-BICs with finite but high $Q$-factors due to perturbations in various forms \cite{27,28}. The quasi-BICs with high $Q$-factors have been  widely applied in many applications such as biological sensing \cite{29}, nonlinear optical processes \cite{30,31,32,33,34}, and unidirectional emission \cite{35,36,37,38}. In addition, the electromagnetic field enhancement caused by quasi-BIC resonance with high $Q$-factor has significant potential to boost TMOKE. While there have been several attempts to enhance TMOKE by utilizing quasi-BIC resonance supported by those structures consisted of the etched magneto-optical materials \cite{39,40}, the patterning of magneto-optical materials involves complex fabrication processes, which may introduce significant challenges to the experimental implementation. Very recently, a periodic arrangement of silicon nanocuboids, combined with an etchless magneto-optical layer and a metal substrate system, has been proposed to enhance TMOKE through the coupling of the quasi-BIC and the surface lattice resonance modes. The TMOKE amplitude achieved by this scheme at normal incidence reaches 0.42, which is six times greater than the one obtained by pure quasi-BIC only  \cite{41}, however, it is still notably lower than the theoretical maximum value of 2. To achieve more practical and impactful applications, it is crucial to further increase the TMOKE amplitude. 

To this end, we propose an all-dielectric structure to achieve high TMOKE based on etchless bismuth-substituted yttrium iron garnet (BIG) by taking full advantage of quasi-BIC in one-dimensional resonant grating waveguide structure. The waveguide structure is composed of a four-part period grating and a BIG waveguide layer.  The simulation results show that in the proposed one-dimensional resonant grating waveguide structure, the amplitude of TMOKE can reach as large as 1.978 at the quasi-BIC resonance. The results obtained from finite-element method simulations agree well with those of coupled mode theory (CMT). As an exemplary application, we design a refractive index sensor based on the resonantly enhanced TMOKE. Our work not only achieves high TMOKE without the need to etch BIG but also sets an example for enhancing TMOKE at the nanoscale and provides a feasible solution for designing integrated magneto-optical devices such as sensors, magnetic storage devices, and nonreciprocal photonic devices.

\section{MODEL AND THEORETICAL ANALYSIS}
TMOKE is defined as the relative change of the reflected intensity for $p$-polarized light when the transverse magnetization is reversed, which arises from the change in the surface boundary conditions of the magnetic layer induced by magnetic field \cite{42}. The TMOKE magnitude can be expressed by the following formula\cite{43},
\begin{equation}
\mathrm{TMOKE}=2 \frac{R(H+)-R(H-)}{R(H+)+R(H-)},\label{eq 1}
\end{equation}
where $R$($H$+) and $R$($H$-) represent the reflectance in the positive and negative magnetic field directions.

The proposed all-dielectric one-dimensional resonant grating waveguide structure aimed at enhancing TMOKE is illustrated in Fig. 1(a). It is composed of a four-part periodic arrangement of grating on a BIG waveguide layer deposited on a SiO$_2$ substrate. It is assumed that the substrate has semi-infinite thickness. The unit cell is shown in Fig. 1(b). The top layer consists of a four-part periodic grating with a period $p$ = 800 nm and a thickness $h_g$ = 300 nm. For this grating, the first and third parts composed of SiO$_2$ have identical width $d_a$, while the second and fourth parts, filled with air, exhibit different widths as $d_b$ = $d_0$ + $\Delta d$ and $d_c$ = $d_0$ - $\Delta d$, respectively, where $d_0$ denotes their initial width. Here, we set $d_a$ =$f_{H}$$p$= 0.25$p$ and $d_0$ = 0.25$p$. The ratio of $\Delta d$ to $d_0$, represented by $\delta$ = $\Delta d$/$d_0$, is defined as the perturbation parameter. When setting a difference in width between the second and fourth parts within the unit cell, it introduces perturbations into the grating waveguide system. The refractive index of SiO$_2$ is 1.45. The second layer is a BIG waveguide with a thickness $h_{wg}$ = 300 nm. When a direct current magnetic field of $\sim$0.05 T along the \textit{z} direction is applied to the BIG film, the permittivity tensor of BIG in the frequence range of interest can be expressed as \cite{44,45} 
\begin{equation}
\varepsilon=\left(
\begin{matrix}
6.25 & 0.06 i & 0 \\
-0.06 i & 6.25 & 0 \\
0 & 0 & 6.25
\end{matrix}
\right).\label{eq 2}
\end{equation}

To achieve a high $Q$-factor resonance, we utilize quasi-BIC induced by the engineered Brillouin zone folding. The first Brillouin zone (FBZ) for unperturbed and perturbed structure are shown in Fig. 1(c), which are indicated by the solid and dashed black box, respectively. The blue shaded area shows that the FBZ has been reduced by half in the $x$ direction due to the period doubling. To clearly show the origin of the quasi-BIC resonance in the proposed one-dimensional resonant grating waveguide structure, we initially calculate its band structure in the absence of the external magnetic field, i.e., the permittivity of BIG is 6.25. The momentum properties are simulated using the finite element method implemented in the commercial software COMSOL Multiphysics. In the following simulation calculations, the Floquet periodic boundary conditions are adopted to the unit cell, and the perfectly matched layers are utilized in the $y$ direction.  Here, we consider the transverse-magnetic (TM)-like eigenmode in the waveguide layer. When $\delta$ = 0, the four-part periodic grating degenerates to the two-part periodic grating with $p^{\prime}$=$p$/2, and its band structure is represented by the blue line in Fig. 1(d). It is worth noting that the TM-like mode lies below the light cone (gray), behaves as a guided mode, and cannot couple with the incident light owing to wave-vector mismatch. While $\delta$ changes from zero to a nonzero value, the period of grating becomes $p$, and its band (red) results from folding the band of the two-part periodic grating structure (blue). Then, the guided mode rises above the light line, where it couples with incident light and transforms into guided mode resonance (GMR). Since the GMR arises from a small geometric perturbation, the coupling between the guided mode and incident light can be regarded as a quasi-BIC resonance \cite{33,46,47,48,49,50}. Section I in the Supplemental Material is referred to provide a more detailed description of the physical mechanism of the quasi-BIC resonance. 
\begin{figure}
\includegraphics[width=\linewidth]{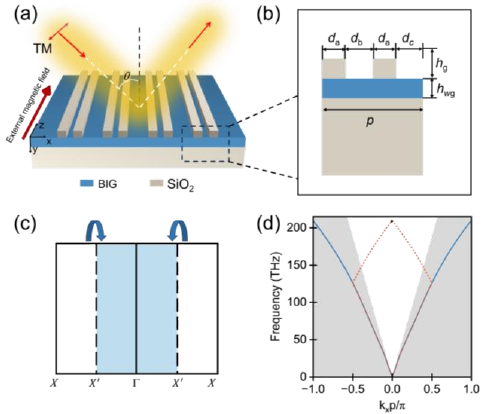}
\caption{(a) The schematic of the compound grating waveguide structure composed of a SiO$_2$ four-part periodic grating layer, an etchless BIG waveguide layer, and a SiO$_2$ substrate. (b) The detailed geometric parameters of a unit cell. (c) The first Brillouin zone of the structure with unperturbed and perturbed cases. (d) The blue line represents the band structure of the two-part periodic grating structure, while the red dashed line represents that of the four-part periodic grating structure. The gray shaded area represents the light cone. }
\end{figure}

To investigate the optical response of the one-dimensional resonant grating waveguide structure, the reflectance enabled by the quasi-BIC resonance is calculated. For illustrative purposes, we set the perturbation parameter $\delta$ to be 0.5 and the incidence angle to be 3°. The dashed lines in Fig. 2(a) depict the simulated reflection spectra under opposite magnetic fields. The Fano resonance peak emerges at 1453.75 nm under a positive magnetic field and at 1454.36 nm under a negative magnetic field, see Fig. 2(a). Owing to the reverse of the direction of the external magnetic field, the dielectric magnetization is rotated, leading to a shift of the resonant wavelength. The intensity modulation of the reflected light enabled by the resonance shift produces S-shaped TMOKE, as indicated by the dashed line in Fig. 2(b). 

To interpret the quasi-BIC resonance enhanced TMOKE, the CMT involving a resonant mode is employed to describe the optical response of the structure. Within this framework, the dynamical equation for the amplitude of the resonant mode can be described by \cite{51}  
\begin{equation}
  \frac{\mathrm{d}a}{\mathrm{d}t} =(j\omega_{0}-\gamma_{1}-\gamma_{2})a+\boldsymbol{\kappa}^{T}\left|\mathrm{~s}_{+}\right\rangle,\label{eq 3}
\end{equation}
\begin{equation}
\left|\mathrm{s}_{-}\right\rangle=C\left|\mathrm{~s}_{+}\right\rangle+ \boldsymbol{D}a,\label{eq 4}
\end{equation}
where $a$ corresponds to the resonance amplitude, $\omega_{0}$ is the resonant frequency, assuming that the resonance decays into two ports with decay rates $\gamma_{1}$ and $\gamma_{2}$ , respectively, and  $\gamma=\gamma_{1}+\gamma_{2}$  is the total decay rate. $\left|\mathrm{~s}_{+}\right\rangle=\left(\mathrm{~s}_{1+} \quad \mathrm{~s}_{2+} \right) ^{T}$  and  $\left|\mathrm{~s}_{-}\right\rangle=\left(\mathrm{~s}_{1-} \quad \mathrm{~s}_{2-} \right) ^{T}$ are the vectors representing the inputting and outgoing waves. $\boldsymbol{\kappa}=\left(\kappa_{1} \quad \kappa_{2} \right)^{T}$ and $\boldsymbol{D}=\left(d_{1} \quad d_{2} \right) ^{T}$ represent the input and output coupling matrix, respectively. According to energy conservation, $\kappa_{i}$ and $d_{i}$ can be expressed as $\sqrt{2\gamma_{i}}e^{j\theta_{\kappa_{i}}}$ and $\sqrt{2\gamma_{i}}e^{j\theta_{d_{i}}}$, respectively. $C$ is the scattering matrix for the direct transmission and reflection and can be described by
\begin{equation}
C= {e^{j\varphi}}\left( 
\begin{matrix}
r_{b} & jt_{b}\\
jt_{b} & r_{b} 
\end{matrix}
\right),\label{eq 5}
\end{equation}
where the phase $\varphi$ depends on the choice of the reference plane, and the real numbers $r_{b}$ and $t_{b}$ correspond to the direct reflection and transmission coefficients, respectively. The reflection coefficient $r$ can be represented as follows based on Eqs. (3), (4), and (5):
\begin{equation}
r=\frac{s_{1-}}{s_{1+}}=e^{j\varphi} r_b+\frac{d_1 \kappa_1}{j\left(\omega-\omega_{0}\right)+\left(\gamma_{1}+\gamma_{2}\right)} ,\label{eq 6}
\end{equation}

The elements in Eqs. (3) and (4) must satisfy certain conditions to ensure energy conservation. The restrictions are as follows:
\begin{equation}
C^{\dagger} C=I ,
\label{eq 7}
\end{equation}

\begin{equation}
\boldsymbol{D}^{\dagger} \boldsymbol{D}=\boldsymbol{\kappa}^{\dagger} \boldsymbol{\kappa}=2 \gamma ,
\label{eq 8}
\end{equation}
\begin{equation}
C \boldsymbol{\kappa}^{*}+\boldsymbol{D}=0,
\label{eq 9}
\end{equation}
\begin{equation}
C^{T} \boldsymbol{D}^{*}+\boldsymbol{\kappa}=0 ,
\label{eq 10}
\end{equation}
From Eq. (5) and Eq. (9), we have
\begin{equation}
\kappa_{2}^{*}=-\frac{1}{j t_b}\left(r_{b} \kappa_{1}^{*}+d_{1} e^{-j \varphi}\right),\label{eq 11}
\end{equation}
Taking the complex conjugate of Eq. (11)
\begin{equation}
\kappa_{2}=\frac{1}{j t_b}\left(r_{b} \kappa_{1}+d_{1}^{*} e^{j \varphi}\right),\label{eq 12}
\end{equation}
Substituting Eq. (11) and Eq. (12) into Eq. (8), we can obtain
\begin{equation}
2 \gamma_{2}=\frac{1}{t_{b}^2}\left(2 r_b^2 \gamma_1+r_b \kappa_1^* d_1^* e^{j \varphi}+r_b \kappa_1 d_1 e^{-j \varphi}+2 \gamma_1\right),\label{eq 13}
\end{equation}
Rearranging Eq. (13), we can get
\begin{equation}
\frac{\gamma_2}{\gamma_1}=1+2\left(\frac{r_b}{t_b}\right)^2+2\left(\frac{r_b}{t_b^2}\right) \cos (\phi),\label{eq 14}
\end{equation}
where the phase difference between the reflection coefficient of the background and the reflection coefficient of the resonance can be defined as $\phi$ = $\varphi$ – ($\theta_{\kappa_{1}}$+$\theta_{d_{1}}$). Furthermore, the reflection spectra can be calculated by
\begin{widetext}
\begin{eqnarray}
R=\frac{\left[r_{b}\left(\omega-\omega_0\right) \pm \sqrt{2 \gamma_1^2+2 \gamma_2^2-r_{b}^2\left(\gamma_1+\gamma_2\right)^2-\frac{1}{r_{b}^2}\left(\gamma_1-\gamma_2\right)^2}\right]^2+\frac{1}{r_{b}^2}\left(\gamma_1-\gamma_2\right)^2}{\left(\omega-\omega_0\right)^2+\left(\gamma_1+\gamma_2\right)^2}.\label{eq 15}
\end{eqnarray}
\end{widetext}
The values of $r_{b}$, $t_{b}$, and $\varphi$ are obtained by calculating the real and imaginary parts of the reflection coefficient of the proposed structure without perturbation. The value of $\theta_{\kappa_{1}}$+$\theta_{d_{1}}$ can be obtained from the phase of the reflection coefficient at the resonant wavelength. The ratio of $\gamma_2$  and $\gamma_1$  can be calculated using Eq. (14). For a positive magnetic field, the above calculated parameters are: $\gamma_1$ = 0.0153 THz, $\gamma_2$ = 0.0177 THz,  $\omega_0$= 206.3622 THz, $r_{b}$ = 0.073, $t_{b}$ = 0.997, $\varphi$ = 0.704 rad, $\theta_{\kappa_{1}}$+$\theta_{d_{1}}$ = 0.768 rad. For a negative magnetic field, the calculated parameters are:  $\gamma_1$ = 0.0151 THz, $\gamma_2$  = 0.0175 THz, $\omega_0$ = 206.276 THz, $r_{b}$ = 0.073, $t_{b}$= 0.997, $\varphi$ = 0.717 rad, $\theta_{\kappa_{1}}$+$\theta_{d_{1}}$ = 0.794 rad. The solid lines in Fig. 2(a) show the reflection spectrum under opposite magnetic fields, calculated using Eq. (12). The TMOKE corresponding to Fig. 2(a) is depicted in Fig. 2(b), exhibiting a sharp Fano-like shape. The TMOKE spectrum consists of the peak and dip with sign change. It is notable that the amplitude of the TMOKE can reach up to 1.578. The results obtained by the simulation are in agreement with the method of CMT, as shown in Fig. 2.

\begin{figure}
\includegraphics[width=\linewidth]{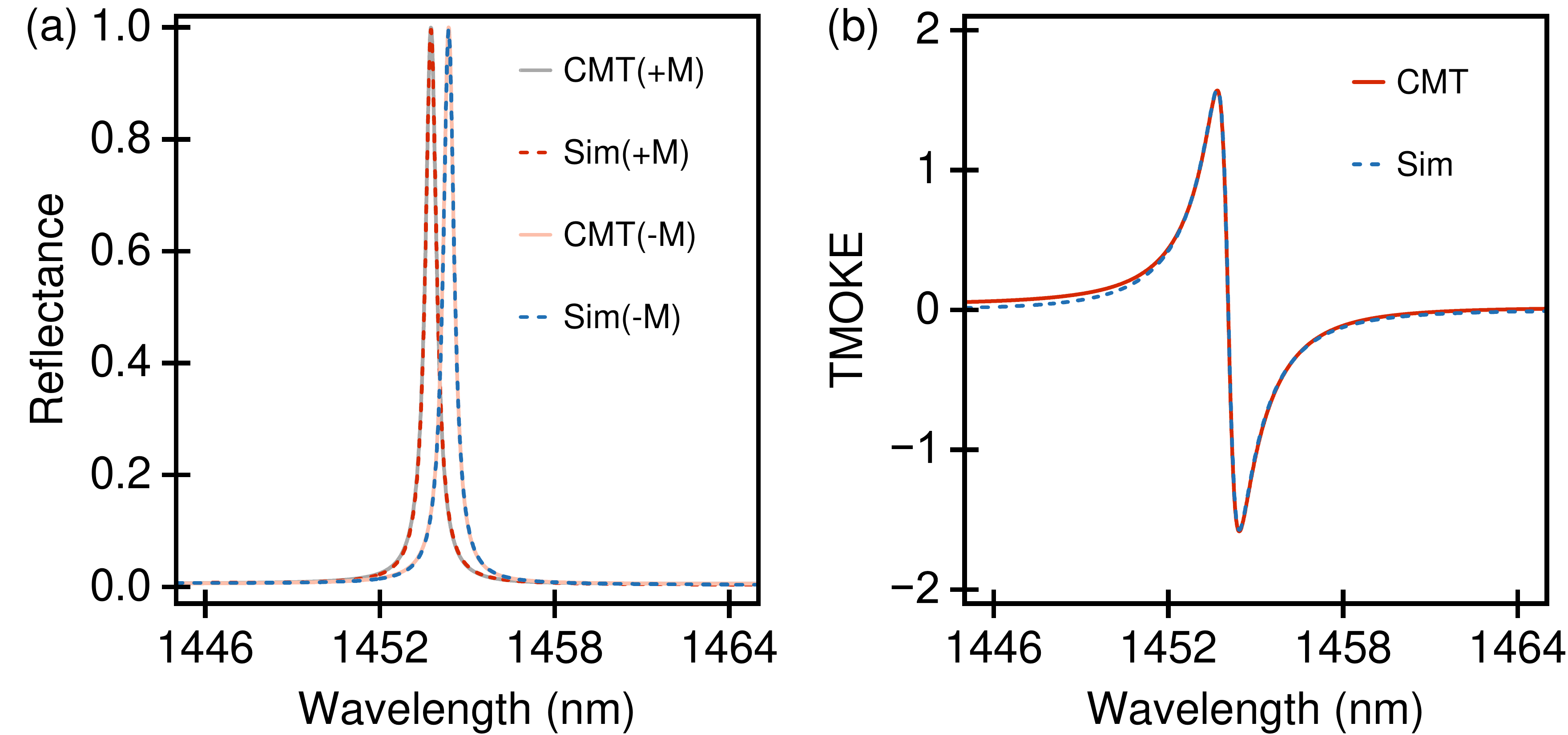}
\caption{The reflection spectra and TMOKE as functions of wavelength are depicted in (a) and (b), respectively, for an incident angle of $\theta$ = 3° and $\delta$ = 0.5.  }
\end{figure}

\section{Quasi-BIC resonantly enhanced TMOKE}
Now we investigate the effect of the perturbation parameter $\delta$ on the reflection and the magnitude of TMOKE. The reflectance spectra of the resonant grating waveguide structure with two opposite directions of material magnetization and different $\delta$ at a fixed oblique angle $\theta$ = 3° for TM polarization, are given in Fig. 3(a). The red line indicates the reflectance under a positive magnetic field, while the blue line indicates the reflectance under a negative magnetic field. The insets show the magnetic field distributions ($\lvert{H_z}\rvert$) at the resonance wavelengths. To quantify the field-enhancement effect, the incident magnetic field intensity is normalized. As the parameter $\delta$ gradually increases from nearly zero to unity, the resonance width of the reflectance peak significantly is broadened due to the enhanced coupling between the leaky guided mode and the positive first-order evanescent diffraction field. Meanwhile, the reflectance peak slightly shifts toward shorter wavelengths due to the changes in the refractive index distribution within the unit cell as $\delta$ varies \cite{52}. When $\delta$ = 1, the Fano resonance peak appears at 1453.04 nm under a positive magnetic field, and at 1453.65 nm under a negative magnetic field. Owing to the reverse of the direction of the external magnetic field, the dielectric magnetization is rotated, leading to a shift of the resonant wavelength. Empowered by the GMR effect, the magnetic field at quasi-BIC is strongly localized inside the BIG waveguide layer, exhibiting a significant enhancement of approximately 50 times compared to the incident magnetic field. When $\delta$ = 0.5, the Fano resonance peak emerges at 1453.75 nm under a positive magnetic field and at 1454.36 nm under a negative magnetic field, with further enhancement in the magnetic field intensity. When $\delta$ = 0, the four-part periodic grating degenerates to a two-part periodic grating, causing the Fano resonance width to vanish, which corresponds to the guided mode. We also analyze the dependence of the $Q$ factor of the quasi-BIC on the perturbation parameter $\delta$. As $\delta$ approaches zero, the $Q$ factor increases rapidly; however, as $\delta$ increases to 1, the $Q$ factor decreases sharply. Given that the $Q$ factor is directly related to the ability to confine magnetic field, the magnetic energy inside the BIG waveguide layer exhibits a similar downward trend. (see detailed information in Section II of the Supplemental Material). The intensity modulation of the reflected light enabled by the resonance shift produces S-shaped TMOKE, as given by Fig. 3(b). When $\delta$ = 1, the amplitude of TMOKE is 0.676. When $\delta$ is reduced to 0.5, the amplitude of TMOKE increases further to 1.578. Notably, when $\delta$ continues to be reduced to 0.25, the amplitude of TMOKE can reach up to as large as 1.946. This is because as the perturbation parameter $\delta$ gradually decreases from unity to near zero, the width of the resonance narrows significantly, and the field distribution becomes increasingly localized. Meanwhile, the field intensity is notably enhanced within the waveguide. The localization of the field defines light-matter interactions, and thereby directly affects the magnitude of the magneto-optical effect. In fact, according to the qualification in the expression of TMOKE, shifts in the narrowing resonant peaks can give rise to a higher TMOKE compared to the wide resonant peaks. 
\begin{figure}
\includegraphics[width=\linewidth]{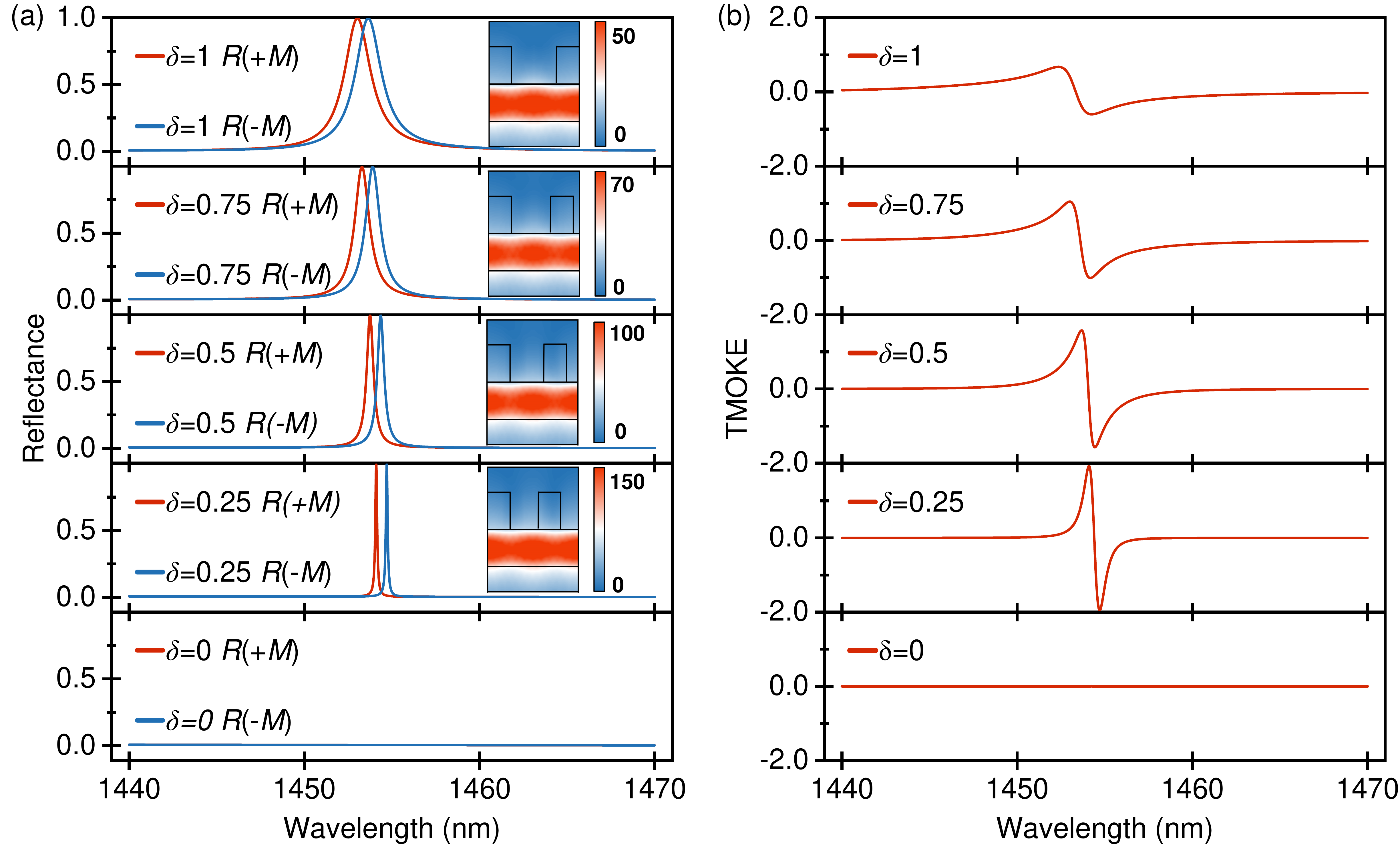}
\caption{Effects of the perturbation parameter $\delta$ on optical and magneto-optical responses: (a) Reflection of the structure as a function of wavelength for a fixed oblique angle $\theta$ = 3° under opposite magnetic fields. The insets depict the distributions of $\lvert{H_z}\rvert$ at the reflectance peaks. The corresponding TMOKE is depicted in (b).  }
\end{figure}
\begin{figure}
\includegraphics[width=8cm]{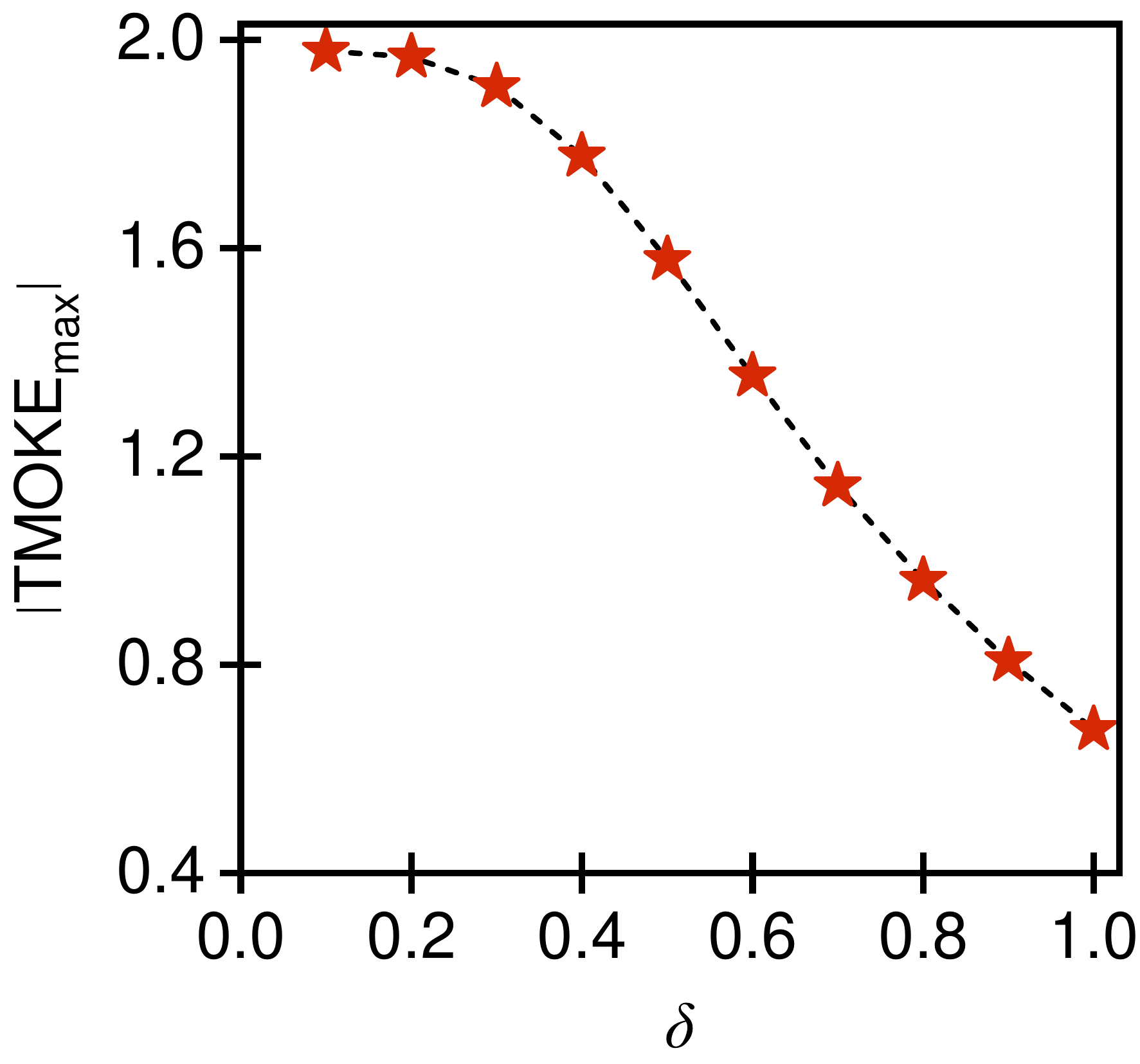}
\caption{The $\lvert\text{TMOKE}$$_\text{max}\rvert$ as a function of perturbation parameter $\delta$.}
\end{figure}

From the analysis above, it can be found that reducing the perturbation parameter $\delta$ leads to larger TMOKE amplitudes. Therefore, we further calculate the effect of perturbation parameter $\delta$ on   (the maximum of $\lvert\text{TMOKE}\rvert $  value) in the designed structure, as shown in Fig. 4. As the perturbation parameter $\delta$ decreases from 1 to 0.1, the TMOKE amplitude increases from 0.676 to 1.978. Initially, the TMOKE amplitude increases rapidly as $\delta$ decreases. However, as $\delta$ approaches approximately 0.2, the rate of increase begins to slow down, and the amplitude change becomes more gradual. If the perturbation parameter $\delta$ is further reduced, it can get very close to the theoretical maximum value of 2.

According to fundamental symmetry arguments, the TMOKE activity is related to the angle of incidence of light \cite{53}. We fix the perturbation parameter $\delta$ at 0.5 and calculate the reflectance and TMOKE spectra as functions of the incident angle and wavelength. The reflectance of the resonant grating waveguide structure as a function of the angle of incidence, is depicted in Fig. 5(a), it is evident that the resonant wavelength redshifts with the increase of the incident angle, which is due to the approximately direct proportional relationship between the resonant wavelength and the angle of incidence for a fixed grating constant. The corresponding TMOKE is depicted in Fig. 5(b). Away from the resonances, the TMOKE is minimal. However, at the resonance, the pronounced positive (red) and negative (blue) peaks are observed, with the amplitude of TMOKE reaching up to 1.63. In addition, for each pair of incident angles that are equal in magnitude but opposite in sign, the TMOKE spectra exhibit opposite signs. For normal incidence as shown in Fig. 5(b), the TMOKE disappears due to the symmetry of the structure.
\begin{figure}
\includegraphics[width=\linewidth]{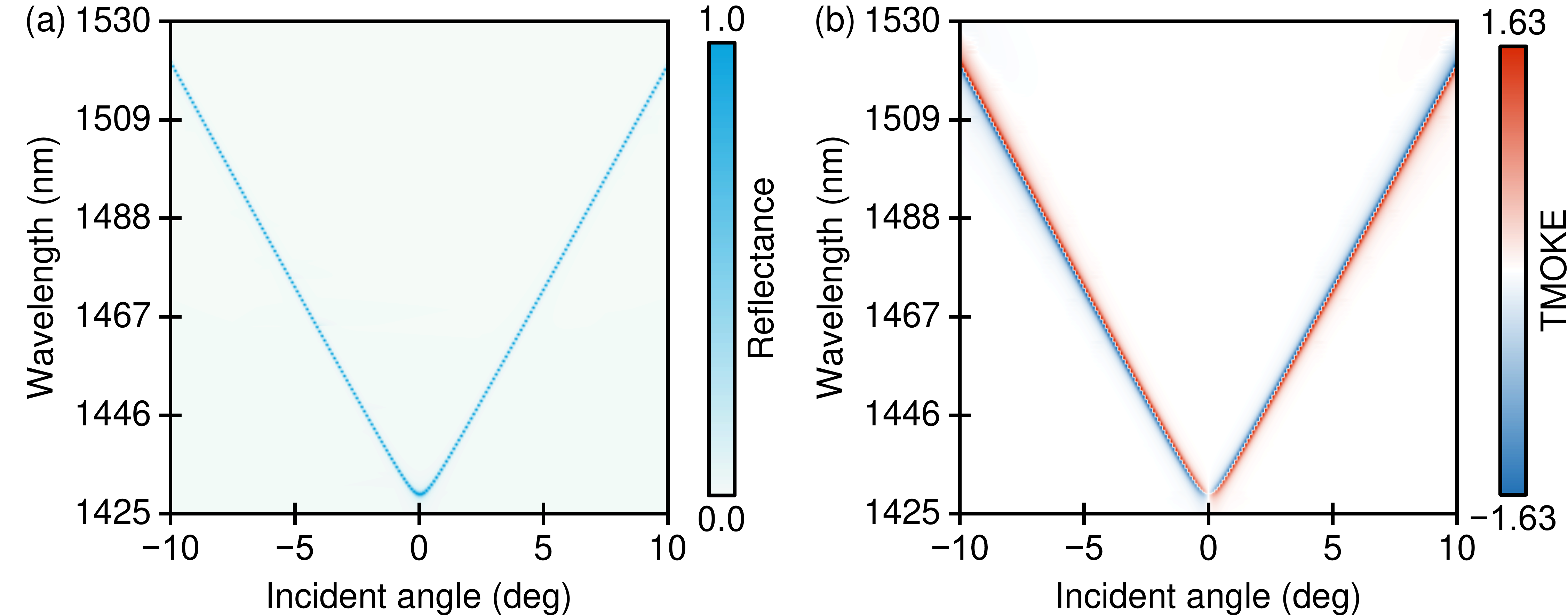}
\caption{Numerical results for (a) reflection and (b) TMOKE as functions of wavelength (horizontal axis) and angle of incidence (vertical axis).  }
\end{figure}

\section{Refractive index sensing application}
Due to the significant enhancement of light-matter interaction provided by the quasi-BIC resonance, the TMOKE in the proposed structure can be further utilized for refractive index sensing to detect minute changes in analyte. Without loss of generality, the geometric parameters are aligned with those in Fig. 3, and $\delta$ = 0.5. As shown in Fig. 6(a), when the refractive index $n_a$  increases gradually from 1.0 to 1.2 in increments of 0.05, the optical responses shift toward longer wavelengths. The TMOKE corresponding to Fig. 6(a) is depicted in Fig. 6(b), all spectra display Fano-like peaks, with positions exhibiting a distinct redshift sensitive to changes in the refractive index. The refractive index sensitivity is defined as: 
$S$=$\lvert \frac{\Delta \lambda}{\Delta n_a} \rvert$, where $\Delta \lambda$ represents the wavelength shift of the Fano-like resonance, and $\Delta n_a$ denotes the change in the refractive index of the analyte. The unit of $S$ is nm/RIU$^{-1}$. As shown in Fig. 6(c), the resonant wavelength varies with the change of the refractive index of the analyte, exhibiting a linear relationship with a slope of 110.66. Therefore, the sensitivity we obtained is $S$ = 110.66 nm/RIU, which is higher than the previous value \cite{54}. A significant factor in evaluating sensor resolution is the figure of merit (FOM), which is calculated as the ratio of sensitivity ($S$) to the line width ($\Gamma^{\prime}$ ) of the TMOKE curve (for each  $ n_a$). 
Given that TMOKE exhibits a Fano-like shape, we can fit the numerical results in Fig. 6(b) using the following function to obtain $\Gamma^{\prime}$ [9].
\begin{equation}
	\mathrm{TMOKE}=A+B \frac{\left(\frac{q \Gamma^{\prime}}{2}+\lambda-\lambda_{\text {res }}\right)^2}{\left(\frac{\Gamma^{\prime}}{2}\right)^2+\left(\lambda-\lambda_{\text {res }}\right)^2},
	\label{eq16}
\end{equation}
where the fitting constants A and B represent the background and the overall peak height, respectively, $q$ is the Fano parameter, and $\lambda_{\text {res }}$ represents the wavelength corresponding to the lowest position of the TMOKE spectrum. The FOM values of the resonant grating waveguide structure are shown in Fig. 6(d), with a corresponding maximum FOM of 299.3 RIU$^{-1}$. As $ n_a$  increases, the FOM value slightly increases, indicating that $\Gamma^{\prime}$ decreases with the increase of $ n_a$ . 
\begin{figure}
\includegraphics[width=\linewidth]{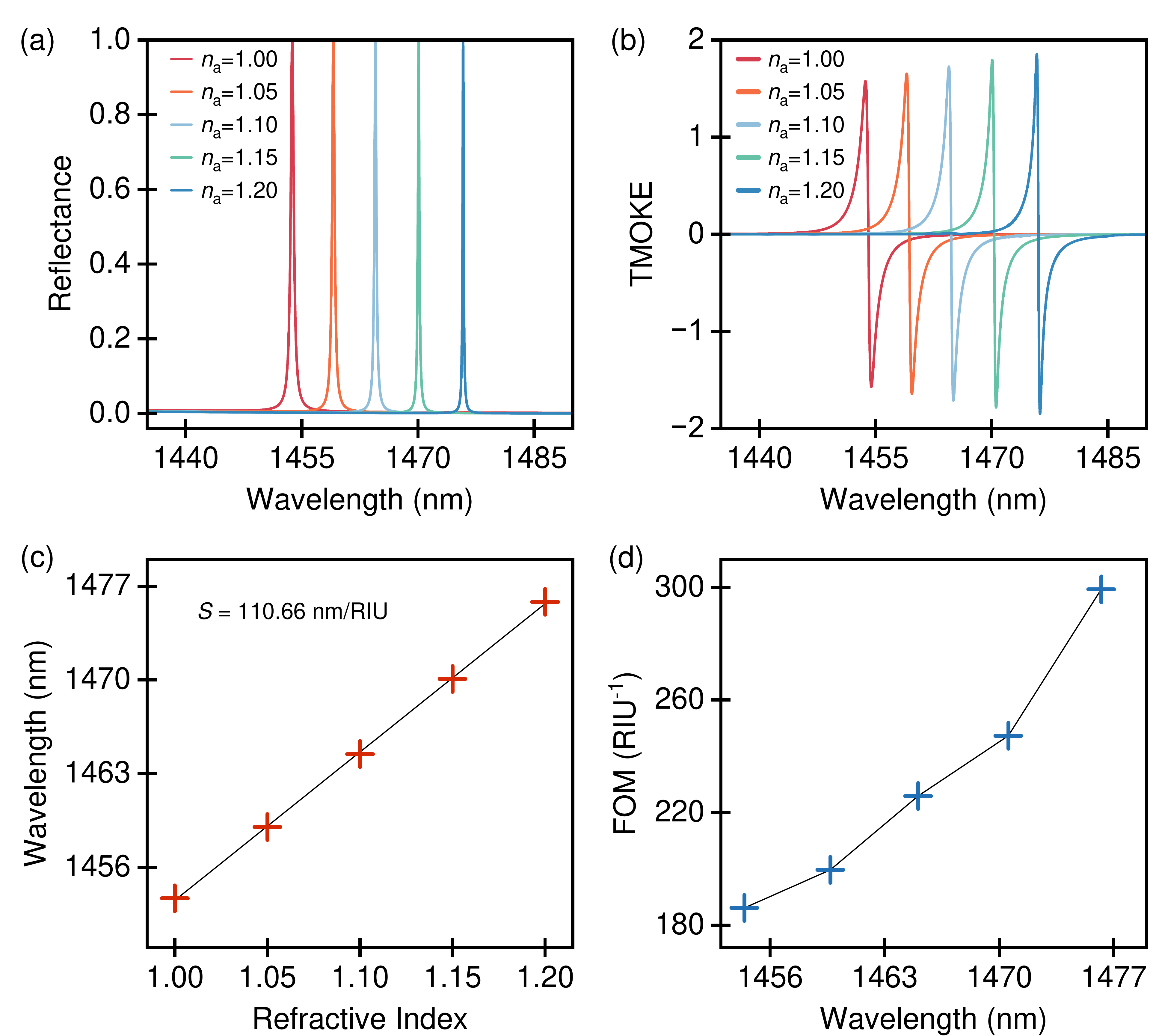}
\caption{(a) The optical response and (b) TMOKE of the proposed sensor as a function of the refractive index of the analyte. (c) The wavelength of the maximum TMOKE value, which varies with the refractive index of the analyte, with the red dashed line showing a linear fit to this dataset. (d) The figure of merit values as a function of the wavelengths corresponding to the lowest position of the TMOKE spectrum.  }
\end{figure}

\section{CONCLUSIONS}
In conclusion, we propose a straightforward method to enhance the TMOKE based on an all-dielectric one-dimensional resonant grating waveguide structure composed of a four-part periodic grating layer and a waveguide layer. Assisted by the quasi-BIC, the TMOKE can be greatly enhanced due to the resonance with an ultrahigh $Q$-factor. Band structure analysis indicates that the emergence of this quasi-BIC stems from the band folding of the band edge to the first Brillouin zone $\Gamma$ point. The numerical result based on the finite element method shows that the amplitude of the TMOKE can reach up to 1.978 at the quasi-BIC resonance, which is very close to the theoretical maximum values of  2. The results obtained from CMT and finite-element method simulations agree very well. The TMOKE can be further tuned by varying $Q$ factor through adjustments to the perturbation parameter, revealing the tunability of the quasi-BIC resonance as well as the amplitude of the TMOKE in the wavelength of interest. In addition, increasing the refractive index of the analyte results in a slight increase in the amplitudes of the TMOKE. The sensitivity and corresponding FOM of the nanostructure sensor are 110.66 nm/RIU and 299.3 RIU$^{-1}$, respectively. Our work not only achieves high TMOKE without the need to etch BIG but also offers insights into the design of efficient TMOKE photonic devices, paving the way for potential applications in various fields, including non-reciprocal photonic devices and magnetic storage devices.

\begin{acknowledgments}
This work was supported by the National Natural Science Foundation of China (Grants No. 12064025, No. 12304420, No. 12264028, No. 12364045 No. 12364049), the Natural Science Foundation of Jiangxi Province (Grants No. 20212ACB202006, No. 20232BAB201040, and No. 20232BAB211025), the Young Elite Scientists Sponsorship Program by JXAST (Grants No. 2023QT11 and No. 2025QT04).
\end{acknowledgments}

\nocite{*}


%

\end{document}